\documentstyle[psfig,amstex,amssymb]{mn}
\begin{document}

\title{Surface Brightness Measurements for APM Galaxies}

\author[Z. Shao, S. Maddox, B. Jones, P. Coles]
       {Z. Shao\footnotemark $^{1,2}$, S. J. Maddox$^{2}$, J. B. Jones$^{2}$ and P. Coles$^{2}$ \\
        $^{1}$Shanghai Astronomical Observatory, Chinese Academy of Science, 
Shanghai 200030, P.R. China\\
        $^{2}$School of Physics \& Astronomy, University of Nottingham,
               Nottingham NG7 2RD, UK \\
        }
\date{Accepted 28 August 2002} 

\pagerange{\pageref{firstpage}--\pageref{lastpage}} \pubyear{0000}

\maketitle

\label{firstpage}

\begin{abstract}
This paper considers some simple surface brightness (SB) estimates
for galaxies in the Automated Plate Measuring Machine (APM)
catalogue in order to derive homogeneous SB data for a very large
sample of faint galaxies. The isophotal magnitude and area are
used to estimate the central surface brightness and total
magnitude based on the assumption of an exponential SB profile.
The surface brightness measurements are corrected for field
effects on each UK Schmidt  plate and the zero-point of
each plate is adjusted to give a uniform sample of SB and total
magnitude estimates over the whole survey. Results are obtained
for 2.4 million galaxies with blue photographic magnitudes
brighter than b$_J$ = 20.5 covering 4300 deg$^2$ in the region of
the south galactic cap. Almost all galaxies in our sample have
central surface brightness in the range 20 to 24 b$_J$ mag
arcsec$^{-2}$. The SB measurements we obtain are compared to
previous SB measurements and we find an acceptable level of error
of $\pm 0.2$ b$_J$ mag arcsec$^{-2}$. The distribution of SB
profiles is considered for different galaxy morphologies for the
bright APM galaxies. We find that early-type galaxies have more
centrally concentrated profiles.

\end{abstract}

\begin{keywords}
surveys - galaxies: photometry - surface brightness.
\end{keywords}

\section{Introduction}

\footnotetext{E-mail: zyshao@@center.shao.ac.cn} 

\label{sec_intro}

Surface brightness (SB) is one of the fundamental parameters
describing a galaxy. It plays an important role in many diverse
aspects of extragalactic astronomy and cosmology, from identifying
the whole family of galaxy populations to modelling their
different distributions and motions in the Universe. These aspects
are generally related to the formation and evolution of galaxies,
as well as to the nature of large scale structures in the
Universe.

Unfortunately, accurate measurements of the SB of galaxies rely
critically on high quality observations with perfect sky
conditions and adequate telescope time, so it is not easy to get
high quality SB data for galaxies, especially for low surface
brightness galaxies (LSBG). Large, homogeneous data samples are
even more difficult to construct, and so uniform SB data for
galaxies is very scarce compared to other parameters, such as
position, colour, redshift and morphology. Hitherto relatively few
sets of SB measurements have been available; the total number of
such galaxies is only a few thousand, and the samples have either
concentrated on LSBGs or been limited only to bright galaxies
(e.g.  de Jong \& van der Kruit 1994; de Jong 1996; Jansen et al.
2000; Heraudeau \& Simien 1996; Lauberts \& Valentijn  1989; Impey
et al. 1996; Morshidi-Esslinger et al. 1999, hereafter MDS).

On the other hand it is not  difficult to make a rough estimate
of a galaxy's SB using  parameters related to both the luminosity
and size of a galaxy.  These two kinds of basic parameters, such
as magnitude and effective radius of galaxies, are common in many
catalogues.  One can consequently obtain a larger sample of SB
data of galaxies at the expense of  some accuracy for individual
objects. The statistical analysis of a large number of galaxies
will not sensitively depend on the accuracy of the SB measurement
for each individual galaxy because reasonable errors in SB can be
allowed for in the analysis.

The APM catalogue includes both the magnitude and isophotal size
for each galaxy so these can be used to estimate the SB of
galaxies. Based on this strategy, some work has already used SB
estimates from the APM survey. C\^{o}t\'e et al.  (1999) used the
mean isophotal SB of APM galaxies to select the candidates for an
HI survey of LSBGs. More recently Cross et al.  (2001) calculated
the effective SB of galaxies in the 2dF Galaxy Redshift Survey and
estimated their bivariate brightness distribution, as well as the
number and luminosity density of galaxies.

In this paper, we consider a variety of SB estimates, and present
a method for determining the central surface brightness, $\mu_0$,
of galaxies in the APM Galaxy Survey.  We apply a field correction
across each survey plate, and use a matching procedure over the
whole survey area to yield a homogeneous SB sample of over 2 million
galaxies with a wide magnitude range over a large
volume of space. The results will be used in future papers to
study the distribution of galaxies as a function of surface
brightness, in particular their clustering properties.
We describe the basic data in Section~\ref{sec_apmcat}.
The principles behind the method are described in
Section~\ref{sec measurement}, together with an introduction of
the raw APM parameters that relate to the SB of galaxies. The
field correction and matching process are briefly outlined in
Section~\ref{sec_corrections}, where we also discuss the
uncertainty of our SB estimates and compare to other SB
measurements. In Section~\ref{sec_discussion} we consider the
distributions of $\mu_0$ and the SB profiles of bright APM
galaxies. Finally we summarize our results in
Section~\ref{sec_summary}.

\section{The APM Catalogue}
\label{sec_apmcat}

The APM Galaxy Survey is hitherto the largest galaxy survey used for
the investigation of the large-scale distribution of galaxies. 
It contains over 3 million galaxy positions and apparent magnitudes,
compiled by scanning 269 UK Schmidt Telescope (UKST) survey plates
over the southern sky with the Automatic Plate Measuring (APM)
machine.  A subset of 2.4 million galaxies in 185 UKST fields south of
declination $-20^{\circ}$, compiled from photographic data of a
consistently high quality and representing the first stage of the APM
survey, was described by Maddox et al. (1990a,b). This subset covered
the south galactic cap and is defined by the boundaries of UKST plates
whose centres have galactic latitude between $b = -30^{\circ}$ and
$-40^{\circ}$ (the exact limit depending on star number densities, and
therefore on longitude), as well as declination. It is this subset of
185 plates, representing an area of 4300 square degrees, that we use
in this paper.

The APM identified images using a connected-pixel algorithm, selecting
groups of 16 or more pixels having intensities above a fixed threshold
optical density.  With a mean pixel size of 0.51~arcsec (the plate
scale varies slightly from plate to plate and across individual
plates), this corresponds to a minimum area of 4.3~arcsec$^2$.  The
detection threshold $t$ was set for each plate to be about twice the
{\it rms} noise in the sky, leading to a mean threshold of 12 APM
density units, corresponding to $\sim25\;b_J\;\mbox{mag arcsec}^{-2}$
above the sky background.  The threshold varies within each plate due
to vignetting and emulsion effects and also varies from plate to
plate, because of the different photometric zero-points for each
plate.  The standard deviation in the threshold due to the variation
in photometric zero points from plate to plate is $0.3\;b_J$ mag
arcsec$^{-2}$.

The images were classified as stars, galaxies, blended images and
noise based on comparisons with the median stellar profile as a function
of magnitude, as described in detail by Maddox et al. (1990a).  In
our present analysis we consider only images classified as
galaxies. 
The total magnitude was estimated from the isophotal magnitude and area
of each detected image, assuming a Gaussian light profile. This
process was optimized for faint images near the magnitude limit where
galaxy profiles are significantly affected by seeing. Different plates
have different limiting magnitudes, and so to ensure a uniform limit,
the final sample was further constrained by the magnitude limit of the
poorest plate, 20.5 b$_J$.  Rejecting images fainter than 20.5 b$_J$
in total magnitude produced the final APM Catalogue that we consider
in this paper.

\section{Measurement of Surface Brightness}
\label{sec measurement}

\subsection{APM data}
\label{sec_apm} The APM Catalogue includes measurements of the
isophotal flux and the image area within the detection isophote
for each image. These parameters are used here to estimate the SB
of each galaxy assuming an exponential light profile. The APM
parameters are calculated using emulsion density $D$, rather than
incident flux $f$ measurements. Below the emulsion saturation
$D_{\rm max}$, it is fairly well approximated by a linear
relationship,
\begin{equation}
  D = \left\lbrace
 \begin{array}{lcl}
  \alpha f + \beta & $for$ & f<(D_{\rm max}-\beta)/\alpha \\ D_{\rm max} &
  $for$ & f>(D_{\rm max}-\beta)/\alpha
 \end{array} \right.
\label{eq_density}
\end{equation}
where $\alpha$ and $\beta$ are constants for any particular part
of the plate (Maddox et al. 1990a,b, Cawson  et al. 1987), and the
value of $D_{\rm max}$ is found to vary by $ 2 \% $ across the
field (Maddox et al. 1990b). The incident flux $f$, is the sum of
the sky flux $f_{\rm sky}$, and the object flux $f_{\rm obj}$. So,
the APM measures the sky-subtracted density $D_{\rm obj}$ for
particular pixels of objects, as
\begin{equation}
\begin{array}{rcl}
D_{\rm obj} & = & D-D_{\rm sky}\\
        & = & \alpha f + \beta - (\alpha f_{\rm sky} + \beta )\\
        & = & \alpha f_{\rm obj}
\label{eq_dobj}
\end{array}
\end{equation}
where $D_{\rm sky}= \alpha f_{\rm sky} + \beta $ is the local sky
value which was obtained from a pre-scan and smoothing process
(Maddox, 1988). Obviously, the upper limit of $D_{\rm obj}$ is,
$D_{\rm sat} = D_{\rm max} - D_{\rm sky}$, for any particular
position in a plate. The saturation density $D_{\rm sat}$
corresponds typically to a magnitude surface brightness $\simeq
22.3 \; b_J \; \mbox{mag} \; \mbox{arcsec}^{-2}$ above the sky,
but this figure varies with the sky background level.
Note that this saturation is quite severe: for a typical disk galaxy
with a central SB of $ 21.4 \; b_J \; \mbox{mag} \;
\mbox{arcsec}^{-2}$, the profile is saturated inside 0.8 disk
scale-lengths.

For each image, the APM measures the relative brightness by
summing $D_{\rm obj}$ for pixels above the threshold $t$ ( $\simeq
25$ mag arcsec$^{-2}$ ),
\begin{equation}
  I_{\rm iso} = \sum_{D_{\rm obj}\geq t} D_{\rm obj}.
\label{eq_iiso}
\end{equation}
The threshold $t$ is constant across the field of each plate, but
varies from plate to plate according to the fluctuation of the sky
density for individual plates (Maddox  et al. 1990a,b). When
working with APM data it is convenient to define an `isophotal APM
magnitude', $m_{\rm iso}$, of an image as
\begin{equation}
  m_{\rm iso} = + \: 2.5\log_{10} I_{\rm iso}
\label{eq_miso}
\end{equation}
This APM parameter is related to the actual isophotal magnitude,
$b_{\rm iso}$, which is given by
\begin{equation}
  b_{\rm iso} = Z - 2.5\log_{10} I_{\rm iso}
\label{eq_biso}
\end{equation}
where $Z$ is the magnitude zero-point of the photographic plate.
Also, the isophotal size of the image is given by
\begin{equation}
  A_{\rm iso}=\sum_{D_{\rm obj}\geq t} 1
\label{eq_aiso}
\end{equation}
$I_{\rm iso}$, $t$ and $A_{\rm iso}$ are the three main APM
parameters that we use to measure the SB profile of galaxies. In
former work on APM galaxy survey, Maddox et al.  (1990b) used
these parameters together with the assumption of Gaussian galaxy
profiles to calculate a total APM magnitude $m_{\rm tot}$ which
was then calibrated to $b_{J}$ using CCD magnitudes. For faint
galaxies near the magnitude limit of the APM survey this is a
reasonable approximation, because the galaxy images are small and
the profile shape is dominated by the seeing disc. But for
brighter galaxies an exponential profile is a better approximation
(see Section~\ref{sec_discussion}). In the following subsections,
we mainly use these three fundamental APM parameters to
investigate the SB of galaxies.

Additionally, there are two other  APM parameters that are related to the
profile of galaxies.  Both of them are density weighted parameters.
One is $\sigma^{2}$, defined as
\begin{equation}
  \sigma^{2}= \frac{\sum_{D_{\rm obj}\geq t} (x^{2}+y^{2})D_{\rm obj}(x,y)}{I_{\rm iso}}
\label{eq_sig2}
\end{equation}
where, $D_{\rm obj}(x,y)$ is the density at location $x$ and $y$
relative to the image centroid. Theoretically, $\sigma^{2}$,
together with $A_{\rm iso}$ can constrain the SB profile of the
image, and judge whether it is best fit by a Gaussian, or
exponential, or even an $r^{1/4}$ profile. Unfortunately, in
practice this is not possible, because $\sigma^{2}$ changes by
only about $10\%$ between these three profiles, and this is the
same order as the observational error in $\sigma^{2}$ (see
Section~\ref{sec_discussion}). Additionally, $\sigma^{2}$ can be
seriously affected by other factors, such as the ellipticity and
substructure of the image, and any other asymmetry of the profile
which will tend to increase $\sigma^{2}$. This means that the
measurement of $\sigma^{2}$ is not as robust as $A_{\rm iso}$ or
$I_{\rm iso}$, and so we do not use $\sigma^{2}$ in the
measurement of SB in this paper. In Section~\ref{sec_profiles}, we
consider the distribution of $\sigma^{2}$ for bright galaxies
which have a large image and hence a smaller error in
$\sigma^{2}$, and show that an exponential profile is actually a
reasonable approximation to the SB profile of most galaxies.

Another useful parameter is the ellipticity  of the image, which
is derived from density-weighted second moments. This is can be
easily converted to the apparent axial ratio of the APM image
(Maddox et al.  1990a; Lambas et al. 1992),
\begin{equation}
 e=\left( \frac{b}{a} \right)_{\rm apparent}.
\label{eq_e}
\end{equation}
On the assumption that the areal profiles of an APM image have similar
shapes, then $e$ is independent of other profile parameters, and will
not affect the measurement of SB.  However, $e$ will affect the
estimation of $\sigma^{2}$ as discussed in
Section~\ref{sec_profiles}.

\begin{figure}

\centerline{\hbox{\psfig{figure=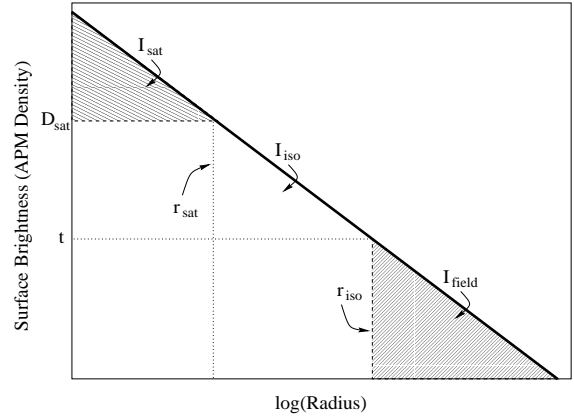,width=0.42\textwidth,angle=-90}}}

\caption{Schematic representation of various quantities used in
our estimation of $\mu_0$. The solid line represents the galaxy SB
profile as a function of radius. $D_{\rm sat}$ is the maximum
measured APM density, and $r_{\rm sat}$ is the radius where this
density is reached. The shaded area above $D_{\rm sat}$ represents
the flux lost due to the saturation, $I_{\rm sat}$.  The isophotal
measurement threshold is shown by the horizontal dotted line,
labeled $t$, and $r_{\rm iso}$ is the isophotal radius. The shaded
area to the right of $r_{\rm iso}$ represents the flux lost due to
the isophotal threshold, $I_{\rm field}$.  The unshaded area below
the solid line is the observed isophotal flux, $I_{\rm iso}$.
\label{fig_iplot} }
\end{figure}
\begin{figure*}

\centerline{
\hbox{\psfig{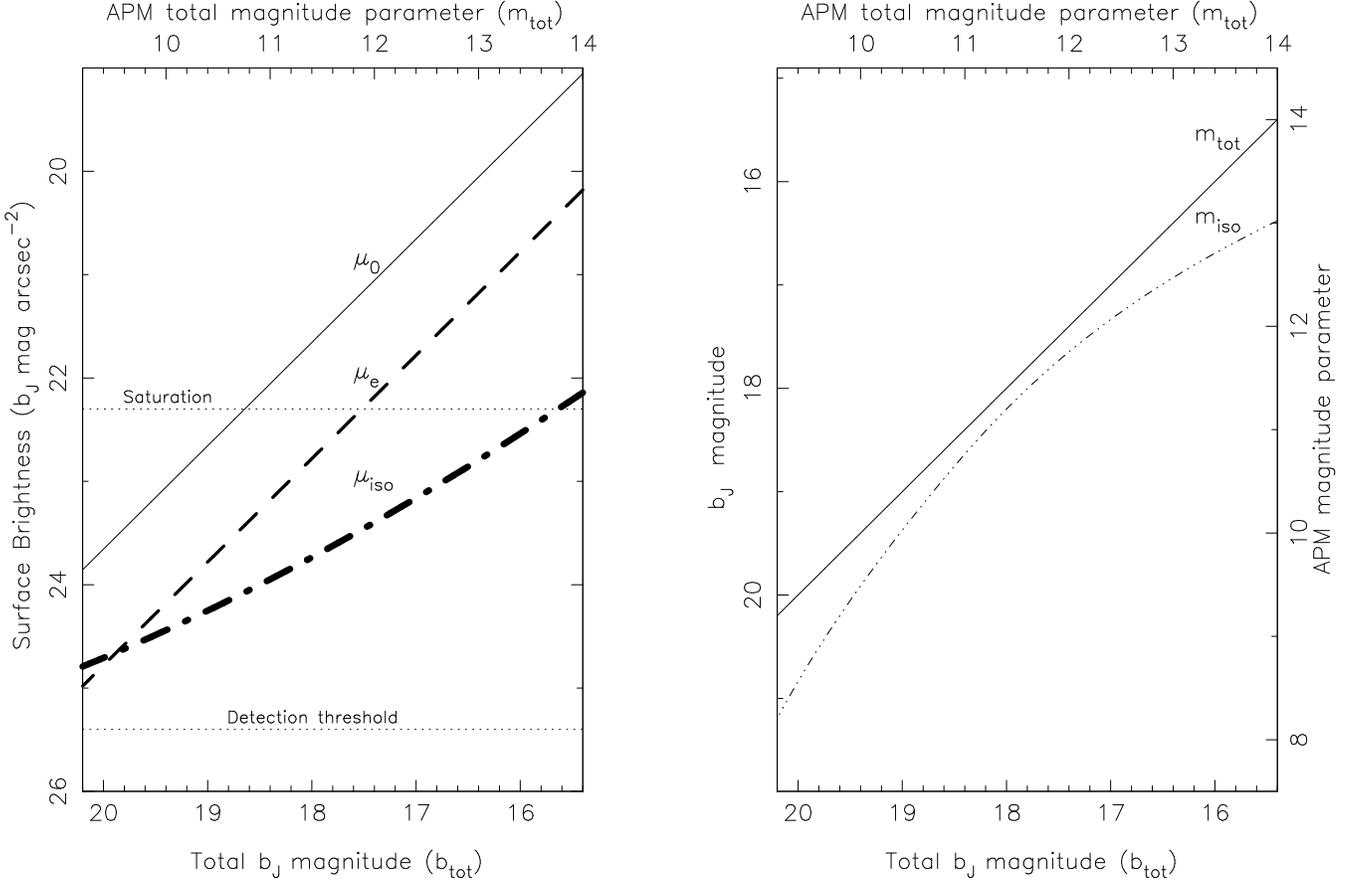}}
}
\caption{ (a) {\em Left:} Simulated surface brightness estimates,
$\mu_{0}$ (central SB), $\mu_{\rm e}$ (mean SB within $r_{\rm e}$)
and $\mu_{\rm iso}$ (mean SB within the detection isophote,
eq.~(\ref{eq_mu25})) for an isophote $t=25.4$ mag arcsec$^{-2}$ as a
function of total magnitude, $b_{\rm tot}$, for
circularly-symmetric exponential profile galaxies. The value of
$r_{0}$ is fixed in this example at 2.15 arcsec (a typical value
for the scale length), and the threshold and saturation value are
set as 12 and 200 APM density units per pixel above the sky level.
(b) {\em Right:} The isophotal magnitude, $m_{\rm iso}$, (dashed
line) as a function of total magnitude, $m_{\rm tot}$, for
circularly-symmetric exponential profile galaxies with $r_{0}$~=
2.15~arcsec, illustrating the effects of the isophotal threshold
and saturation. The total magnitude relation is shown (as a solid
line) for comparison. Results are shown for both APM magnitude
parameters (equations (\ref{eq_miso}) and (\ref{eq_mtot2})) and
for conventional magnitudes.   \label{fig_u0ue} \vspace*{5mm} }

\end{figure*}

\subsection{Modelling the SB profile of galaxies}

If the image of a galaxy is represented by a circularly-symmetric
pure exponential surface brightness profile,
\begin{equation}
  p(r)=p_{0}\exp(-r/r_{0}).
\label{eq_pr}
\end{equation}
The central peak value $p_{0}$, and the characteristic size
$r_{0}$ fully describe the image. In this case, $r_{0}$ is the
exponential scale length.  The total luminosity of the image,
$I_{\rm tot}$, is simply the integral of $p(r)$. The integration
can be divided into three parts which contribute to $I_{\rm tot}$,
\begin{equation}
  I_{\rm tot}=I_{\rm sat}+I_{\rm iso}+I_{\rm field},
\label{eq_itot}
\end{equation}
where $I_{\rm sat}$, $I_{\rm iso}$ and $I_{\rm field}$ represent
the flux lost due to saturation, the observed flux and the
unobserved flux beyond the isophotal size of the galaxy
respectively. These can be written as:
\begin{equation}
  I_{\rm sat} = 2\pi \int^{r_{\rm sat}}_{0}p(r) r dr - 2\pi r_{\rm sat}^{2}D_{\rm
  sat};
\label{eq_isat}
\end{equation}
\begin{equation}
  I_{\rm iso} = 2\pi \int^{r_{\rm iso}}_{r_{\rm sat}} p(r) r dr + 2\pi r_{\rm sat}^{2}D_{\rm
  sat};
\label{eq_iiso2}
\end{equation}
and
\begin{equation}
  I_{\rm field} = 2\pi \int^{\infty}_{r_{\rm iso}} p(r) r dr,
\label{eq_ifield}
\end{equation}
where
\begin{equation}
 r_{\rm iso}=\sqrt{A_{\rm iso}/\pi}=r_{0}\ln(p_{0}/t)
\label{eq_riso}
\end{equation}
denotes the radius corresponding to the isophotal area for
circular images and
\begin{equation}
r_{\rm sat}=r_{0}\ln(p_{0}/D_{\rm sat}) \label{eq_rsat}
\end{equation}
denotes the radius corresponding to the saturated area for
circular images. These quantities are illustrated schematically in
Figure~\ref{fig_iplot}. Of these three parts, $I_{\rm iso}$ is the
only directly observed quantity, and is given by $\sum_{D_{\rm
obj}\geq t} D_{\rm obj}$ (eq.~\ref{eq_iiso}). The measured values
of $I_{\rm iso}$ and $A_{\rm iso}$ can be used together with
eq.(\ref{eq_riso}) and eq.(\ref{eq_rsat}) to find the parameters
$p_{0}$ and $r_{0}$. The resulting equations are not soluble
analytically, but are simple to solve numerically. The actual
procedure that we adopt is as follows: take an initial value of
($p_{0}'$, $r_{0}'$); calculate the ($A_{\rm iso}'$, $I_{\rm
iso}'$) from eq.(\ref{eq_iiso2}), eq.(\ref{eq_riso}) and
eq.(\ref{eq_rsat}). Then compare with the observational APM
parameters ($A_{\rm iso}$, $I_{\rm iso}$), and calculate a
feedback of ($\Delta p_{0}'$, $\Delta r_{0}'$). Several iterations
will lead to the final stable result of ($p_{0}$, $r_{0}$). The
analysis can be adapted to non-circular images by assuming
concentric elliptical isophotes and using the APM ellipticity.

The effects of seeing can also be included in this iterative
process, but for most galaxies the correction is less than 0.1
mag. Our estimates of $p_0$ are based on $I_{\rm iso}$ and $A_{\rm
iso}$, and for images larger than about 100 pixels these
measurements are not sensitive to the seeing.

\subsection{Surface Brightness}

The pair of parameters ($p_{0}$, $r_{0}$) entirely describes the
exponential profile of circularly-symmetric images, so they can be
used to infer the value of any other characteristic SB parameter
for such a galaxy. In particular we can define the central SB in
units of magnitudes per unit area,
\begin{equation}
  \mu_{0}=Z-2.5\log_{10}(p_{0})
\label{eq_mu0}
\end{equation}
where $Z$ is the zero-point for that particular UKST plate. We can
also define an effective SB,
\begin{equation}
  \mu_{\rm e}=\mu_{0}+C
\label{eq_mue}
\end{equation}
where the constant C depends on the exact definition of the
effective SB. It can also be useful to define an effective radius,
$r_{\rm e}$, which is the radius containing half the light of
the galaxy. This is related to the scale size by $r_{\rm e} =
1.678 r_{0}$ for a circularly-symmetric image with an exponential
light profile.  For the SB at $r_{\rm e}$, the constant is
$C=1.822$, and for the mean SB within $r_{\rm e}$, $C =
1.124$.  Also the average SB within the isophotal area (for the
APM scans of UKST blue plates this isophote is $\approx 25$
b$_{J}$ mag arcsec$^{-2}$) can be written as
\begin{equation}
  \mu_{\rm iso}=Z-2.5\log_{10}\left(\frac{I_{\rm iso}}{A_{\rm iso}}\right)
\label{eq_mu25}
\end{equation}
where $A_{\rm iso}$ and $I_{\rm iso}$ are the corresponding
isophotal size and brightness, derived directly  from
eq.(\ref{eq_pr}). Additionally, one can also calculate the `total
APM magnitude' from $I_{\rm tot}$,
\begin{equation}
  m_{\rm tot} =  + \: 2.5\log_{10}(I_{\rm tot})=  + \: 2.5\log_{10}(2\pi
  p_{0}r_{0}^{2}),
\label{eq_mtot2}
\end{equation}
for circular images. This is related to the conventional total
magnitude of the image, which is given by
 \begin{equation}
  b_{\rm tot} =  Z - 2.5\log_{10}(I_{\rm tot}) =  Z - 2.5\log_{10}(2\pi p_{0}r_{0}^{2})
\label{eq_mbtot}
\end{equation}
where $Z$ is the magnitude zero-point.

\begin{figure}
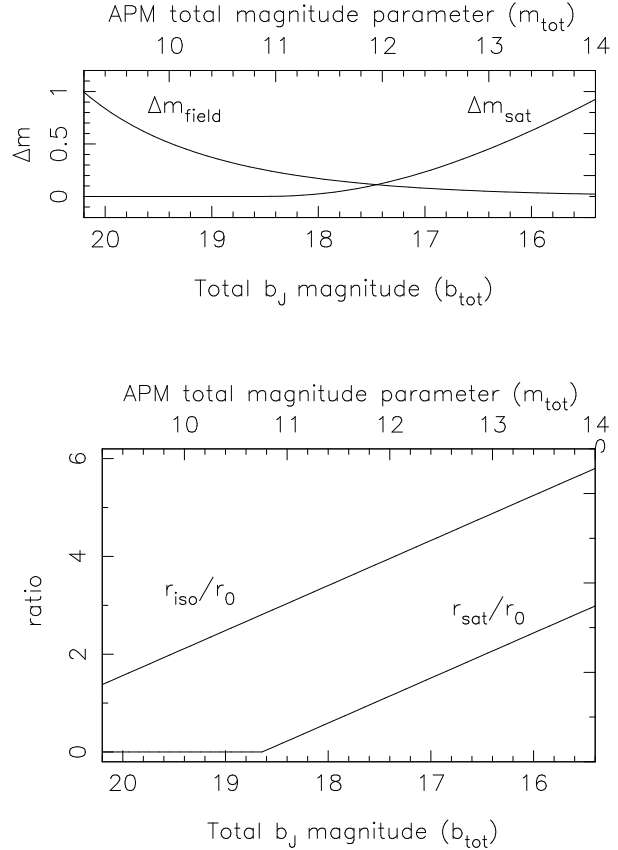

\hbox{\psfig{figure=dm.ps,width=0.45\textwidth,angle=-90}}
\hbox{\psfig{figure=rsat.ps,width=0.45\textwidth,angle=-90}}
\caption{ (a){\it Top:}  The change in total magnitude due to the isophotal
threshold, $\Delta m_{\rm field}$, and saturation, $\Delta m_{\rm sat}$, as a
function of total magnitude.  (b) {\it Bottom:} The ratio of isophotal radius,
$r_{\rm iso}$, and saturation radius, $r_{\rm sat}$, to the scale length,
$r_0$ as a function of total magnitude. The value of
$r_{0}$ is fixed  at 2.15 arcsec, as for Figure 2. 
\label{fig dm}
}
\end{figure}

The relationship between these three SB measures can be seen in
Figure~\ref{fig_u0ue}.  In this Figure, we fix the value of
$r_{0}$ to illustrate the dependencies between these SB parameters,
giving $r_0$ a value of 2.15~arcsec, chosen to be fairly typical of the
APM sample. 
Though the different SB measures are all derived from the same set
of ($p_{0}$, $r_{0}$) data, we see in Figure~\ref{fig_u0ue}a that
some galaxies can show much larger differences between $\mu_{\rm
iso}$ and either $\mu_{0}$ or $\mu_{\rm e}$ than other galaxies.
Figure~\ref{fig_u0ue}b presents the isophotal magnitude, labelled
$m_{\rm iso}$, as a function of the total magnitude for the same
value of $r_0$. This is shown for both the standard magnitude and
for the APM magnitude parameters of equations (\ref{eq_miso}) and
(\ref{eq_mtot2}). $m_{\rm iso}$ can be significantly fainter than
$m_{\rm tot}$, at both high and low SB. At high SB (bright total
magnitude), the flux $I_{\rm sat}$ is lost because of the emulsion
saturation; at low SB (faint total magnitude) the flux $I_{\rm
field}$ is missed because of the isophotal limit.  This also
provides a rough estimate of how much $\mu_{0}$ will be
underestimated if these two parts of missing flux are not
accounted for.
The effect of saturation and isophotal threshold on the magnitude
is shown explicitly in Figure~\ref{fig dm}(a).  As mentioned in
Section~\ref{sec_apm}, bright galaxies may be saturated over a large
fraction of the image.  This can be seen quantitatively in
Figure~\ref{fig dm}(b), which shows the ratio of isophotal radius,
$r_{\rm iso}$, and saturation radius, $r_{\rm sat}$, to the scale length,
$r_0$ as a function of total magnitude.

As described above, this approach only provides a rough
measurement of the SB of galaxies. It does not use the real
profile of different types of galaxies, which may be different
from an exponential disk. It also neglects any internal
structures, such as arms, bars or a central bulge. 
Also, although we have rejected images that are most likely to be
merged pairs, the automated image classification is not perfect, and
there is a residual contamination of about 5-10\% of merged images,
whose profiles will not be well represented by our simple model.

Although for any individual galaxy our SB measurement is unlikely to be
very accurate, it is helpful to constrain the shape of any given
profile and it does allow us to make general comparisons of galaxy
SB within the whole sample. Theoretically, this approach is
suitable for any other profile which is specified by only two
parameters, such as a Gaussian profile,
\begin{equation}
  p(r)=p_{0}\exp\left(-(r/r_{0})^{2}\right)
\label{eq_pgauss}
\end{equation}
or an $r^{1/4}$ law profile,
\begin{equation}
  p(r)=p_{0}\exp\left(-(r/r_{0})^{1/4}\right).
\label{eq_pr4}
\end{equation}
Without any other effective observational constraints on the
profile shape an exponential profile is the most reasonable choice
since it is a good representation for a majority of galaxies. As
discussed in Section~\ref{sec_profiles}, we find that $\sigma^{2}$
as defined in eq.(\ref{eq_sig2}) is not very effective in
distinguishing between different profiles. Additionally, the
exponential profile has been a popular choice in earlier work, so
it allows us to compare with other results. So, in this paper, we
continue our analysis and discussions under the assumption that
galaxies have an exponential SB profile.

Note that these estimates are in terms of the ``APM'' SB, which is
based on  APM density measurements. They are directly related to
true SB only if there is a linear relationship between the real
flux and the APM density (eq.\ref{eq_density}). Clearly the
emulsion saturation violates this assumption, and CCD observations
show that it is not exactly linear below saturation (Cawson et al.
1987). Our fitting technique approximates the saturation as a
simple maximum APM density, and ignores any non-linearity below
the saturation density. This non-linearity will not strongly
influence the ordering of galaxies as a function of their SB, but
more accurate measurements would have to allow for this factor.
Furthermore, if the intrinsic SB is required for a particular
analysis instead of the apparent SB, then the K-correction,
cosmological dimming, and the correction of the apparent
ellipticity of disk galaxies should also be taken into account.

\begin{figure}
\hbox{\psfig{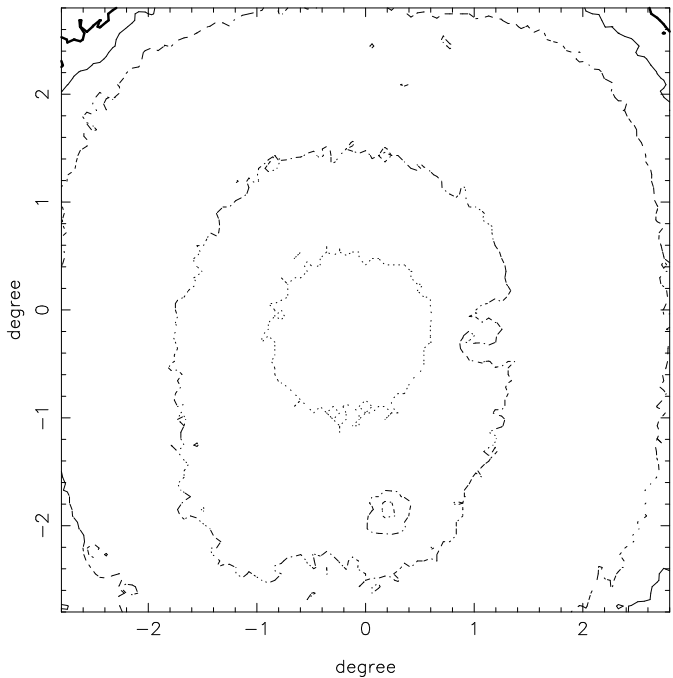}}
\caption{A contour plot of the field correction values, averaged over
185 plates. The dotted line shows $\triangle \mu_{1}=-0.015$ mag
arcsec$^{-2}$ , the dash-dot line 0.0 mag, the dashed line 0.1 mag,
the solid line 0.2 mag and the thick solid line 0.3 mag }

\label{fig vigmap}

\end{figure}

\begin{figure*}
\hbox{\psfig{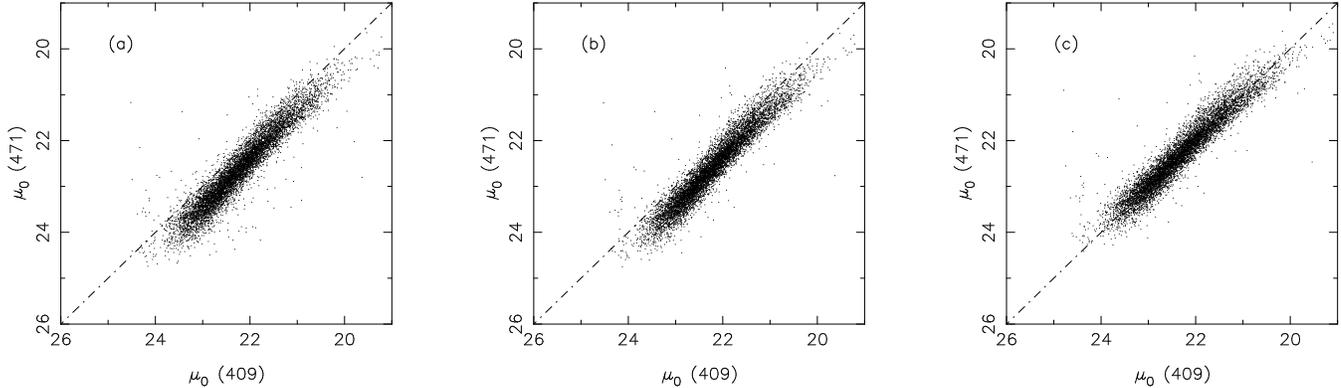}}
\caption{Comparisons of the b$_{J}$ central surface brightness of
galaxies within the overlapping region of UKST fields F409 and
F471 (a) before any correction; (b) after field correction
(Section ~\ref{sec_field}); (c) after further calibration of the
magnitude zero-points by matching results for galaxies in
overlapping regions of UKST plates (as in Section
~\ref{sec_matching}). The {\em rms} scatter in the differences are
in turn 0.27, 0.25, 0.25 mag arcsec$^{-2}$. The dashed line in
each panel shows the line $\mu_0(409)=\mu_0(471)$.
\label{fig_us409471} }
\end{figure*}

\section{Field correction and overall plate matching}
\label{sec_corrections}

To provide a uniform sample of SB measurements over the whole APM
galaxy survey area, we make two further systematic corrections. First,
we correct for the field response function to make the different parts
of each UKST plate consistent from the central region to the edge. Then,
we make an overall photometric zero-point adjustment to each plate, to
give a uniform sample over the whole survey area.

\subsection{Field response corrections}

\label{sec_field}

The variation in the response function across the field of each plate is
caused by two main factors. One is the geometrical vignetting of the
optical system of the telescope (UKST), that reduces the effective
transmission off its axis. A theoretical analysis (Dawe \& Metcalfe
1982) shows the transmission is very flat for images closer than
$2.5^\circ$ from the center of a plate, but quickly falls to $\approx
80 \% $ of the central value at $4^\circ$ off axis.  This reduction is
same for each plate. Another loss in sensitivity is caused by the
differences in the chemical desensitization at different parts of the
UKST survey plates. For most of the survey plates, the original plate
holder has an air gap between the plate and the filter, and the
greater air circulation near the plate edges caused greater
desensitization of the hypersensitized III-aJ emulsion. This effect
causes a multiplicative reduction in the measured flux, which is quite
similar to the geometrical vignetting, but the pattern actually varies
for different plates, because the local weather conditions during
individual observations were significantly different.

Both of these effects can be considered as a change in the
relation between APM densities and true flux.  If two objects have
same flux, $f_{\rm obj}$, but are located at different positions
on a plate, then the geometrical vignetting and the different
desensitization lead to different observed APM densities $D_{\rm
obj}$ via eq.(\ref{eq_dobj}). The constant $\beta$ is essentially
the fog level of the emulsion, which is constant, so the loss in
sensitivity means that the slope $\alpha$  is not a constant
across the plate. It is a function of position in the field, being
reduced by the geometrical vignetting and emulsion desensitization
from centre to the edge. Thus, both these effects can be
considered  as variations in $\alpha$.

Fortunately, $\alpha$ is only a function of position in the field,
and does not depend on the specific properties of an image.
Therefore, the correction process of Maddox et al. (1990b), which
was devised to correct the total magnitude of images, can be
simply applied here to correct the SB of galaxies. According to
eq. (8) of Maddox et al. (1990b), the correction to the total APM
magnitude is
\begin{equation}
  \Delta m_{1} = 2.5\log_{10}\left(\frac{\alpha_{1}}{\alpha_{0}}\right)
\label{eq_deltam}
\end{equation}
where the subscript 0 denotes the centre of a plate and the
subscript 1 denotes the particular position of an object.
According to the relationship between the central SB and the total
APM magnitude of eqs. (\ref{eq_mu0}) and (\ref{eq_mtot2}), we can
see that $\mu_{0}$ will need the same correction as $m_{\rm tot}$:
\begin{equation}
  \Delta \mu_{1} = -
  2.5\log_{10}\left(\frac{\alpha_{1}}{\alpha_{0}}\right).
\label{eq_deltamu}
\end{equation}
The value of $\Delta \mu_{1}$ is determined individually for each
galaxy by estimating the APM sky density in the surrounding 10 arc
minutes.  The apparent variation in sky density is converted to an
effective change in $\alpha$ using an empirical relation which was
chosen to ensure that the number density of galaxies is uniform
when averaged over many UKS plates.  This correction will be
generally applicable for other SB measurements, except for the
mean isophotal surface brightness, $\mu_{\rm iso}$. Since the
threshold $t$ is fixed in terms of observed APM density, the
actual threshold isophote varies over each field and a more
complex correction would be required to ensure that $\mu_{\rm
iso}$ is uniform.

Most survey galaxies are located within the central $3^\circ$
radius of a UKST plate, and so the correction is very small,
typically $\leq 0.1$ magnitude. For galaxies outside this region
the correction can increases up to 0.4 mag. See figure~\ref{fig
vigmap} for the pattern of averaged $\triangle \mu_{1}$ of field
correction. See the figures of Maddox et. al (1990b) for further
details.

Figures~\ref{fig_us409471}(a) to (b) show that the field
correction slightly reduces the {\em rms} difference between the
SB measurements for those galaxies imaged on two independent
plates. F409 and F471 are a pair of contiguous plates with quite a
large overlapping area, and there is a total of 7960 identified
galaxies with measurements from  both the plates. For quite a few
of these galaxies, the distance from the centre of one of the
plates can be large, which means that uncorrected values of
$\Delta \mu_{0}$ can be quite large. Figure~\ref{fig_us409471} (a)
plots the $\mu_{0}$ before field correction and (b) after it. The
{\em rms} scatter in these two plots is reduced from 0.27 to 0.25
mag. This improvement suggests that the {\em rms} $\Delta \mu_{0}$
from the field correction is $\simeq 0.1$.

\subsection{Matching calibrations in plate overlaps over the whole survey}
\label{sec_matching}

Variations in the emulsion sensitivity and observing conditions
mean that each UKST survey plate has a different photometric
zero-point. To produce a uniform set of galaxy SB estimates, we
need to adjust the zero-point of each plate to force a match
between galaxies in overlapping regions. In the APM survey paper
II, Maddox et al. (1990b) described a scheme to match the APM
total magnitudes over the survey. In this paper we apply a similar
procedure to match our SB estimates.

The matching procedure uses the fact that each plate covers an
area of sky that overlaps with its neighbours, so there are many
galaxies that have independent measurements from two neighbouring
plates. For a particular overlap the mean difference between the
two measurements gives an estimate of the difference between the
zero-points of the two plates. Taking all overlaps together we can
find the set of plate zero-points which is most consistent with
these plate-to-plate offsets.

The main difference for this paper is that a galaxy SB profile is
determined by two parameters ($p_{0}$, $r_{0}$) as described in
Section~\ref{sec measurement}, and so both parameters must be
matched. In principle, we could choose to match any two
independent parameters derived from these two, but for simplicity
we choose $\mu_{0}$ and $b_{\rm tot}$.  They are related to
$p_{0}$ and $r_{0}$ through eqs.~(\ref{eq_mu0}) and
(\ref{eq_mbtot}), while the zero-point in eq.~(\ref{eq_mu0})
represents the overall corrected uniform zero-point.  Since we
have assumed exponential profiles in this analysis, the magnitude
zero-points are not the same as the Gaussian total magnitudes that
were applicable to galaxies near the plate limit, as used in
previous work on the APM galaxy survey.

\begin{figure}
\centerline{\hbox{\psfig{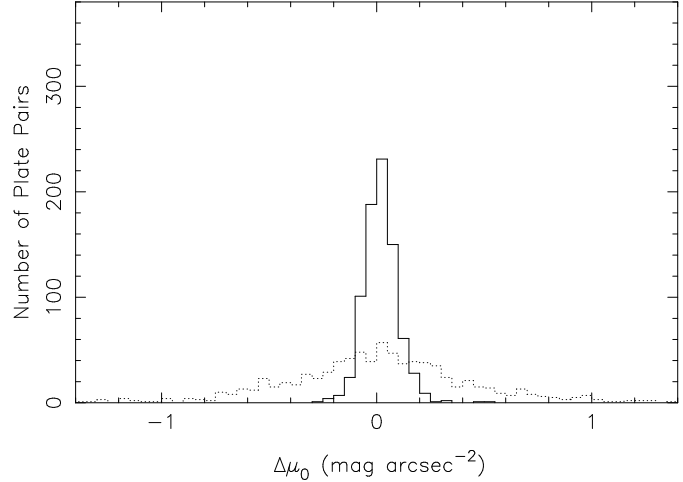}}}
\caption{The frequency distribution of offsets in $\mu_{0}$ for 
galaxies in 809 overlapping fields. The dotted histogram shows
the results before matching, with $\sigma=0.43$  mag
arcsec$^{-2}$; The solid line shows the distribution after
matching, with $\sigma=0.08$ mag arcsec$^{-2}$.
\label{fig_offsetus} }
\end{figure}

\begin{figure}
\centerline{\hbox{\psfig{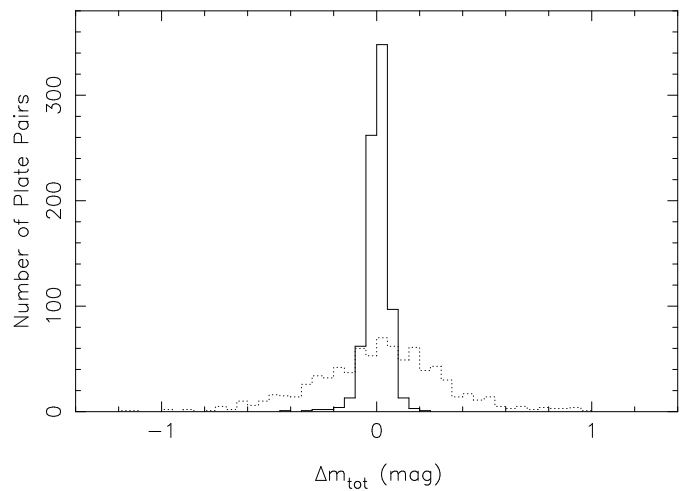}}}
\caption{The frequency distribution of offsets in $m_{\rm tot}$ for
galaxies in 809 overlapping fields. The dotted histogram shows
the results before matching, with $\sigma=0.31$ mag; The solid
line shows the distribution after matching, with $\sigma=0.05$ mag.
\label{fig_offsetms} }
\end{figure}

Images were matched in the overlaps of every pair of plates with
centres closer than $7^\circ.$ For the 269 plates in the south
galactic cap, this gives 809 distinct overlaps with an average of
1405 galaxies brighter than $b_J = 20.5 $ in each overlap.  To
allow for non-linearities which introduce a non-unit slope in the
APM magnitude and SB measurements we used a linear fit to
$(\mu_i-\mu_j)$ {\it vs} $(\mu_i+\mu_j)/2$.
Before matching, the variance in zero-point for the different UKST
survey plates is very large, as can be seen in the offset of
$\mu_{0}$ for galaxies in the overlapping plate pairs. The
distribution of offsets $\bar \mu_1 - \bar \mu_2 $ for all 809
overlaps, is shown in Figure~\ref{fig_offsetus}, and the {\em rms}
scatter is 0.43 mag. After matching, this value is reduced to 0.08
mag. As shown in Figure~\ref{fig_offsetms}, similar reduction in
scatter is seen for $m_{\rm tot}$, where the {\em rms} before
matching is 0.31 mag, and after matching is 0.05 mag.

For different plates, the zero-point corrections for $\mu_{0}$ and
$m_{\rm tot}$ are well correlated, but not exactly the same.  The
scatter about the correlation is caused by the different seeing on
different plates, which leads to a different apparent $r_{0}$. The
matching process has effectively given $r_{0}$ for galaxies on
each individual plate a systematic correction to a common
``zero-point''.

\subsection{Error estimation}
\label{sec_errors}

For an individual galaxy, the deviation of the actual SB profile
from an exponential profile is likely to be the main source of
error in the SB estimate, but this effect is beyond the scope of
this paper. If we assume a given profile, then we can discuss the
``intrinsic'' error in the APM SB caused by the observational
errors in ($I_{\rm iso}$, $A_{\rm iso}$).

The errors in ($I_{\rm iso}$, $A_{\rm iso}$) come from a complex
combination of sources which we consider in two main categories.
First is the observing process, which includes a mixture of
weather conditions and instrumental effects, including the
residual calibration differences between the survey plates that
are not corrected by the field correction function and overall
plate matching. This category also includes photon noise. Second
are the errors from the APM scan itself. These errors can be
estimated directly from repeat scans of selected plates. For
$I_{\rm iso}$ we find this source contributes only about 0.04 mag
uncertainty, and for $A_{\rm iso}$ we find about 5\%. These errors
could be analytically transformed to $\mu_{0}$ following a
standard propagation of errors procedure, but even if the
resulting uncertainty in $\mu_{0}$ is slightly larger, it will be
negligible compared to the observational errors.

\begin{figure}
\centerline{\hbox{\psfig{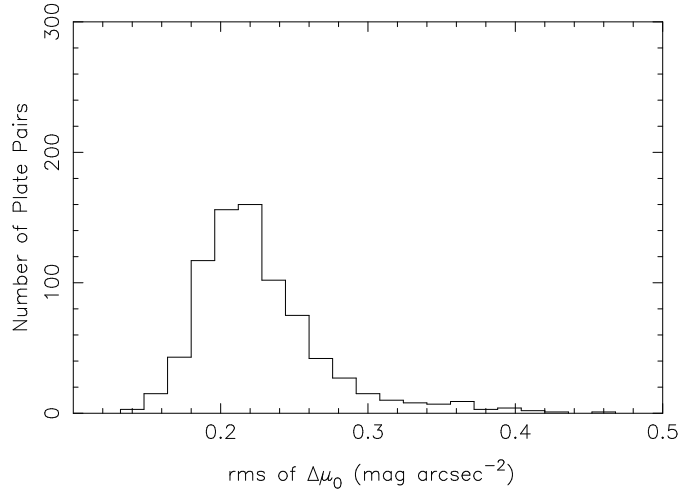}}}
\caption{Frequency distribution of scatters ({\em rms}) of
$\Delta\mu_{0}$ of galaxy pairs for 809 overlapping fields.
\label{fig_rmsdu}
 }
\end{figure}

The overlapping areas of neighbouring plates provide a direct
opportunity to estimate the combined errors in our SB estimates.
Figure~\ref{fig_us409471}(c) shows the residual errors for the
overlapping field of F409 and F471, after applying the field
correction and the overall matching correction, and it can be seen
that there is still an {\em rms} scatter of about 0.25 mag between
the two independent measurements for each galaxy. This implies the
error in each $\mu_{0}$ will be $\simeq 0.18 $ mag.  We have
calculated the {\em rms} scatter for each one of the 809 plate
overlaps, independently, and show the frequency distribution of
these {\em rms} values in Figure~\ref{fig_rmsdu}. For most of
these overlaps, the {\em rms} of $\Delta\mu_{0}$ is less than 0.3
mag, and the mode is at 0.21 mag, corresponding to a typical
uncertainty of 0.15 mag for an individual SB measurement.

If we consider all of the plate overlaps as a whole, the {\em rms}
$\Delta\mu_{0}$ over all 1,282,600 galaxy pairs is 0.23 mag, so
the overall uncertainty for an individual SB measurement has an
{\em rms} of 0.16. This value includes both the random error on
each SB measurement, and the {\em rms} residual difference between
the zero-points for all plates after applying the matching
corrections. The difference between this {\em rms}, and the {\em
rms} within each overlap suggests that the residual error in the
zero-point matching has an {\em rms} of 0.07 mag per plate,
consistent with the overlap residuals.

\begin{table}
\centering \caption{The {\em rms} uncertainty in $\mu_{0}$ for
galaxy pairs with different central surface brightnesses}

\begin{tabular}{|c|r|c|}
\hline\\
 $\mu_{0}$ (mag arcsec$^{-2}$) & no. of galaxy pairs & {\em rms} of
 $\Delta\mu_{0}/\sqrt{2}$ \\
\hline\\
    24.5 $\sim$ 23.5 &  40832 &0.22 \\ 
    24.0 $\sim$ 23.0 & 202094 &0.19 \\ 
    23.5 $\sim$ 22.5 & 480355 &0.17 \\ 
    23.0 $\sim$ 22.0 & 639704 &0.16 \\ 
    22.5 $\sim$ 21.5 & 564329 &0.16 \\ 
    22.0 $\sim$ 21.0 & 373165 &0.16 \\ 
    21.5 $\sim$ 20.5 & 181962 &0.18 \\ 
    21.0 $\sim$ 20.0 &  62253 &0.21 \\ 
    20.5 $\sim$ 19.5 &  14258 &0.23 \\ 
\hline
  \end{tabular}
\end{table}

Additionally, we have considered the error as a function of
$\mu_{0}$. Table 1 lists the {\em rms} difference between the
$\mu_{0}$ measurements in overlaps for a series of $\mu_{0}$
ranges. Galaxies with $\mu_{0}$ about 21.0 $\sim$ 22.0 mag have
the smallest error at 0.16 mag. At both low and high surface
brightness the error increases up to $\simeq$ 0.23.  This tendency
could also be seen in Figure~\ref{fig_us409471}(c). The reason for
the increased error is that $\mu_{0}$ depends on the profile of a
galaxy (see Section~\ref{sec_profiles}), and when we estimate the
total intensity of the galaxy, $I_{\rm tot}$ (eqs.~(\ref{eq_itot})
to (\ref{eq_ifield})), only $I_{\rm iso}$ is directly constrained
by the observational data, while $I_{\rm field}$ and $I_{\rm
sat}$, the other two contributions to $I_{\rm tot}$, involve
extrapolations that are more extreme at low or high SB.

So in summary, the overall ``intrinsic'' uncertainty in $\mu_{0}$
including measurement errors, and zero-point errors is typically 0.2
mag.  Comparing galaxies measured on one plate, the relative
uncertainty is about 0.16 (also see the next subsection).

\subsection{Galaxy surface density and mean $\mu_o$}

\begin{figure}
\centerline{\hbox{\psfig{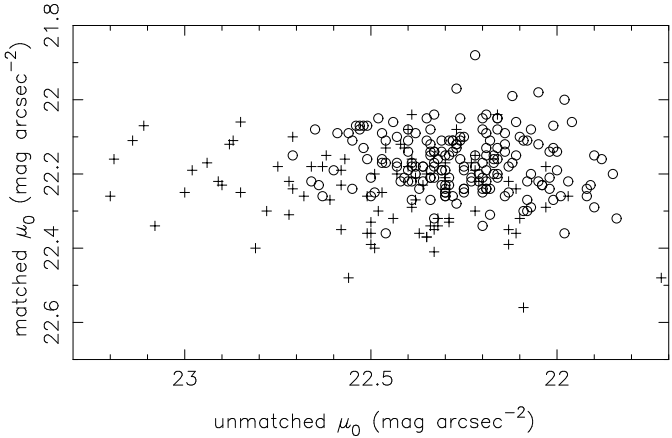}}}
\caption{The distribution of mean $\mu_0$ for all plates in the
survey. Circles show fields where the declination of the centre is
$\le -20^\circ$, and crosses show fields where the declination of the
centre is $>-20^\circ$.
\label{fig_meanmu0}
 }
\end{figure}

After applying the corrections to the individual plate data, we
should be able to define galaxy catalogues selected with uniform
SB criteria. We can make a simple test of the uniformity of the
sample by calculating the mean SB for galaxies on each plate.
Cosmic structure will introduce some intrinsic variation in the
mean SB in different directions on the sky, but this is likely to
be small; indeed the variation in galaxy numbers from plate to
plate at $b_J = 20$ is only $12\%$ (Maddox, Efstathiou \&
Sutherland 1996).  Figure~\ref{fig_meanmu0} shows a plot of the
mean $\mu_0$ for each plate in the survey, before and after
applying the matching corrections. Using all fields, the {\em rms}
scatter before correction is 0.25 mag, and after correction it
reduces to 0.095 mag. However, it is clear that fields centred
with declination  $>-20^\circ$ have a larger scatter,
even after correction. This is partly due to the fact that the
scanned plates with  declination  $>-20^\circ$ are a rather
heterogeneous mix  of 
original plates and non-survey plate copies, and partly due to the
significantly large galactic extinction in this part of sky (see
Schlegel, Finkbeiner \& Davis 1998).  In principle it should be
possible to correct these problems, but for simplicity we have
simply excluded the fields with declination $>-20^\circ$ in the
subsequent analysis.

\begin{figure}

\centerline{(a)\hbox{\psfig{figure=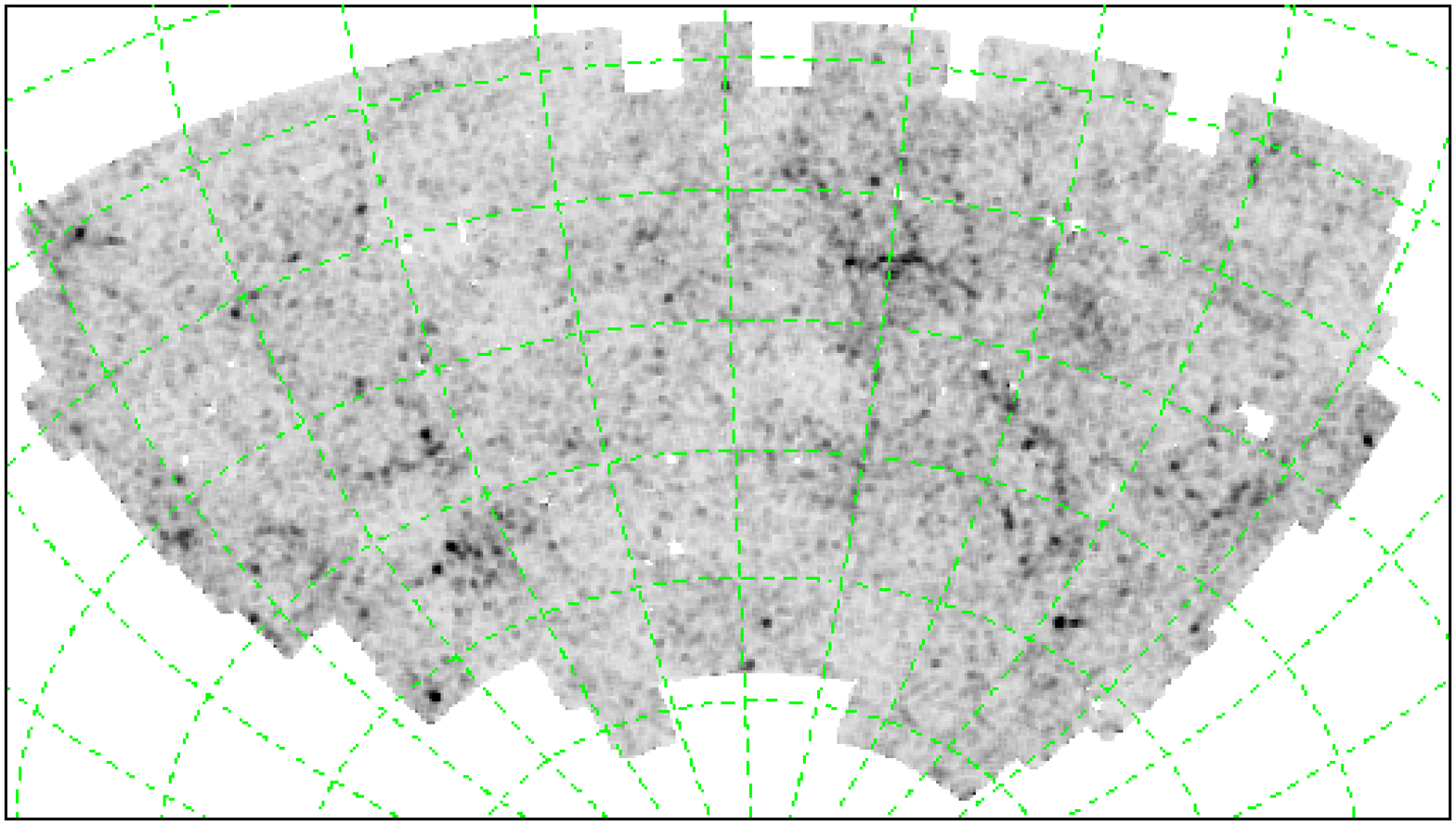,width=0.5\textwidth,angle=-90}}}

\centerline{(b)\hbox{\psfig{figure=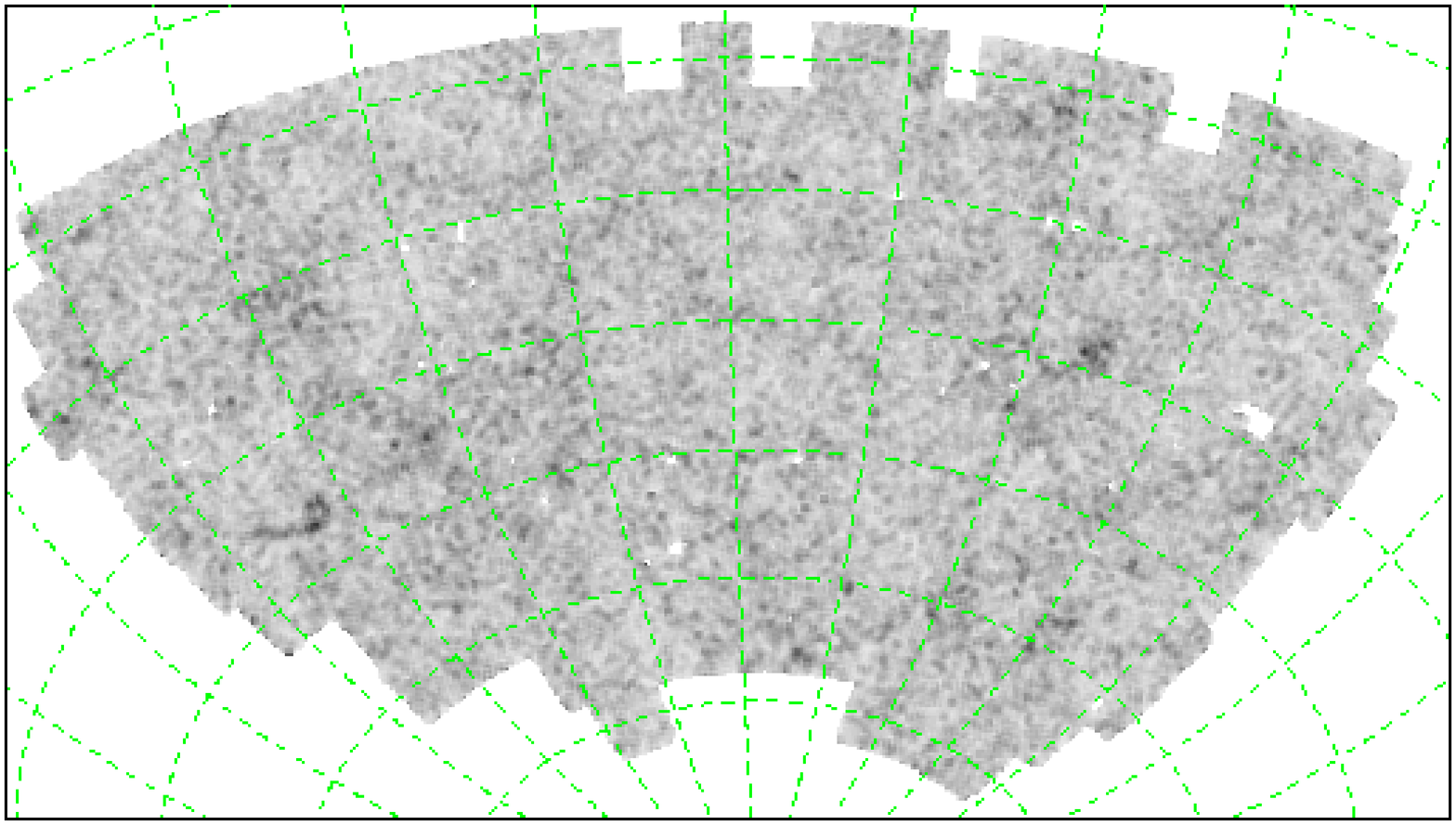,width=0.5\textwidth,angle=-90}}}

\centerline{(c)\hbox{\psfig{figure=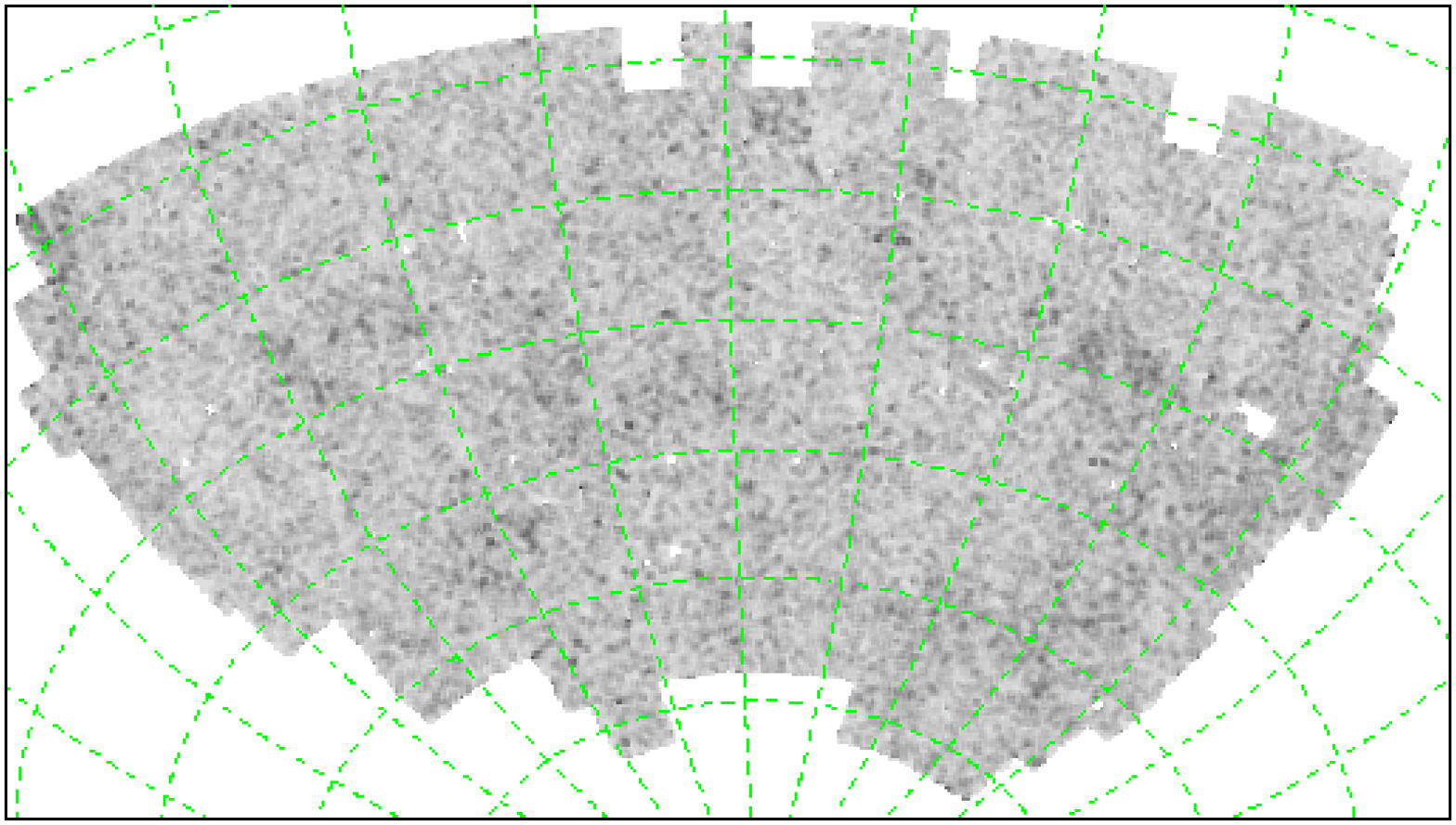,width=0.5\textwidth,angle=-90}}}

\caption{ The distribution of SB selected subsamples of galaxies
projected on the sky: (a) $20.5 < \mu_0 < 21.2 $ $b_{J}$ mag
arcsec$^{-2}$; (b) $22.0 < \mu_0 < 22.2 $ $b_{J}$ mag arcsec$^{-2}$;
(c) $ 23.0 < \mu_0 < 24.5 $ $b_{J}$ mag arcsec$^{-2}$. Each panel
shows a greyscale map shaded according to the surface density of
galaxies in $14'\times 14'$ cells over the SGP. White represents 0
galaxies per cell, and black represents 4 times the mean surface
density.  The dashed lines show right ascension and declination.
\label{fig_skymu}
}
\end{figure}

A further simple check on the uniformity is to compare the average
number of galaxies per plate for SB selected samples.
Figure~\ref{fig_skymu} shows the distribution of galaxies on the
sky for three subsamples selected on SB from high to low SB
respectively. The ranges in $\mu_0$ were chosen to give samples
with roughly the same number of galaxies in each subsample. Note
that the high and low SB samples are selected from the tails of
the SB distribution (see Section~\ref{sec_sb}) and so are most
sensitive to errors in the plate matching.


\begin{figure}

\centerline{\hbox{\psfig{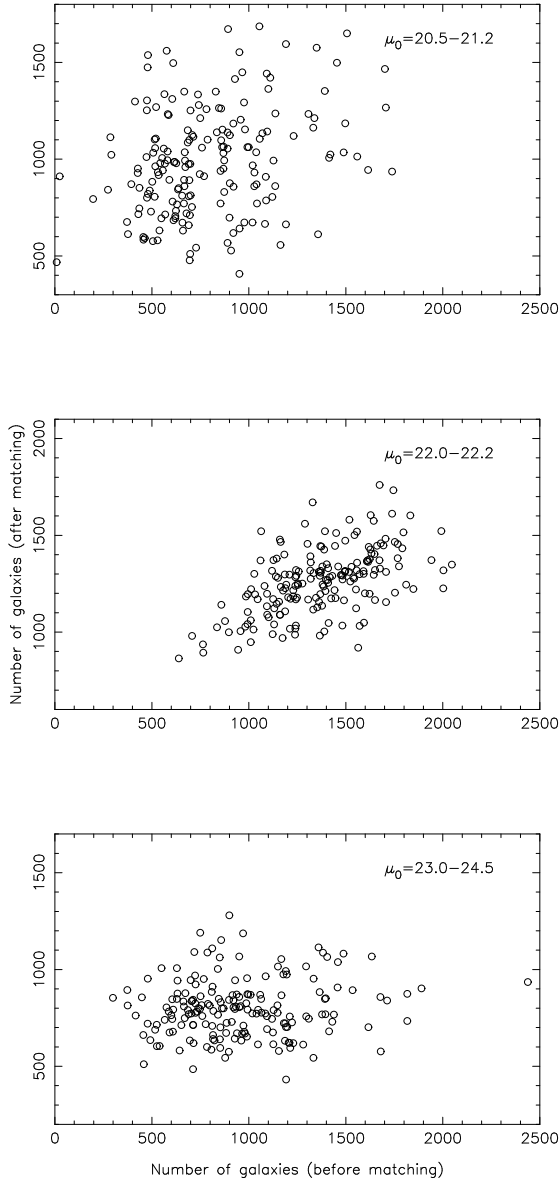}}}
\caption{The number of SB selected subsamples of galaxies per UKST
survey plate's central $5^{\circ}\times 5^{\circ}$ region, before
and after matching correction. Their {\em rms} scatters,
normalized by their average numbers of galaxies, are 0.40, 0.20,
0.35 separately before matching and are reduced to 0.29, 0.13,
0.19 after it. }

\label{fig galno}

\end{figure}

For the 185 plates there are no plates which show a clearly
discrepant number of galaxies, suggesting that our matching
procedure does indeed produce a uniform set of SB measurements.
This improvement can also be seen in figure~\ref{fig galno}, which
plots the numbers of galaxies in each plate for these three
subsamples before and after matching correction. After matching,
the  fractional {\em rms} scatters in the number of galaxies on each
plate are 0.27, 0.13, 0.17 respectively for the three samples. These
are close to what is expected for randomly selected $5^{\circ}\times
5^{\circ} $ patches of sky.  

The higher {\em rms} for the high SB sample reflects the higher
clustering amplitude of high SB galaxies, which is clearly apparent in
Figure~\ref{fig_skymu}. We plan to quantify these differences in in
clustering in a future paper.

\subsection{Checking the APM results and photometric calibration
with deeper CCD data}

It is important to test the accuracy of the APM surface
photometry, particularly the photometric calibration, through a
comparison with well-calibrated higher-resolution CCD data. To
this end, new CCD surface photometry has been performed on faint
($b_J \simeq 18$ to 22 mag) galaxies in the region around the
Hubble Deep Field South (HDF-S), which lies in the area of the
available APM data (at galactic coordinates ($l$,$b$) =
($328^\circ$, $-49^\circ$)).  Much well-calibrated public imaging
data are available around the HDF-S, and ground-based HDF-S CCD
observations cover a sufficient area to provide the numbers of
galaxies required for these tests. The Goddard Space Flight Center
HDF-S public data (Teplitz et al. 2001; Palunas et al. 2000) were
selected for the task because they combine a large image area,
standard broadband photometric passbands and good photometric
calibrations. The Goddard HDF-S data were recorded with the Big
Throughput Camera on the 4-metre Blanco telescope at the
Cerro-Tololo Inter-American Observatory (Wittman et al. 1998) and
cover a 47 arcmin by 46 arcmin area. An independent measurement
found that the seeing of the blue data was 1.7~arcsec in the
central 15 arcmin by 15 arcmin region; although not good, this is
comparable to or better than the photographic APM data. Pixels
were 0.43 arcsec in width.

    We measured the SB profiles of galaxies in this region through the
following steps. Firstly, images were detected in the B-band CCD
data using the SExtractor package (Bertin 1998; Bertin \& Arnouts
1996). This provided image centroid positions, ellipticities and
orientations. For each galaxy image, mean surface brightnesses
were computed in a series of concentric elliptical annuli defined
by these SExtractor image parameters.  Then the surface brightness
was plotted against semi-major axis for each image and an
exponential light profile fitted, defining the range in semi-major
axis for these fits by visual inspection.  Care was taken to avoid
fitting to the core of the image where seeing effects were
significant. This provided a central surface brightness and
exponential scale length for each selected galaxy.  Finally, the
surface brightness results were transformed from the instrumental
blue magnitude system to the photographic $b_J$ system using the
Goddard HDF-S transformation and the standard B to $b_J$ equation
of Blair \& Gilmore (1982). This used colours measured within
identical elliptical apertures in the B and V data. The overall
uncertainty in the individual central surface brightness
measurements is about 0.1~mag.

    An independent check of these results was made using other
public ground-based HDF-S data. SExtractor total magnitudes were
computed using the Anglo-Australian Observatory public
CCD observations for the region immediately around the WFPC2
field. These were transformed to the $b_J$ band and compared
with the total magnitudes under the fitted exponential profiles
derived from the Goddard HDF-S data. Six galaxies with Goddard
photometry were identified
in the AAO field. The difference in the total magnitudes was
$+0.032 \pm 0.030$~mag, confirming the validity of the results
obtained from the Goddard public data. The surface photometry
from the Goddard data can therefore be used to check the
APM photographic results.

    The HDF-S CCD surface photometry gave central surface brightnesses
$\mu_0$ and scale lengths $r_0$ for a large sample of galaxies with
$b_J = 17.5$ to 22 for which the exponential profile was
judged to be an adequate fit. Of this sample, 24 objects
appeared in the APM Galaxy Catalogue. Of these 24, two
galaxies were rejected from further consideration because
their angular sizes were so small (isophotal areas smaller
than 100 pixels) that seeing was likely to affect appreciably
the results from the APM data. This left 22 galaxies in the
region around the HDF-S that had surface photometry both
from the APM data and from the CCD observations. These
galaxies were used to test the reliability of the surface
brightness results derived from the APM data.

\begin{figure}
\centerline{\hbox{\psfig{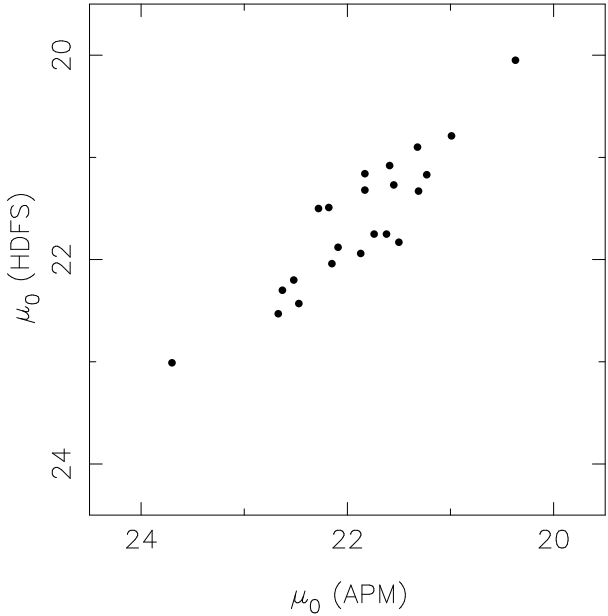}}}
\caption{Comparison of central surface brightness of HDF-S and
this paper. The {\em rms} scatter is 0.29 mag arcsec$^{-2}$.}

 \label{fig hdfs}
\end{figure}

   Figure~\ref{fig hdfs} shows the relationship between the CCD
$\mu_{0}$ and our APM measurements. The {\em rms} scatter is 0.29
mag arcsec$^{-2}$, and the difference in their zero-points is 0.26
mag arcsec$^{-2}$.  This dispersion is larger than expected from
simply adding the uncertainty of 0.18 in the APM measurements in
quadrature with 0.1 from the CCD measurements.  The extra scatter
is probably due to the fact that we have not corrected the
non-linear correspondence between flux and the emulsion density of
the survey plates. Note also that nearly all of the galaxies in this
sample have $\mu_{0}<22$~mag arcsec$^{-2}$, which is brighter than
the saturation level of the UKST emulsion and thus the APM
measurements depend on extrapolated galaxy profiles.  

The scatter could be reduced by restricting the range in radius used
to fit the CCD data so that it matches more closely the range
available in the APM data. However, this would not be a true
representation of the uncertainties in the APM SB estimates.  We want
to compare the APM results with results from CCD surface photometry
that have been obtained using standard surface photometry methods.
Note that there will also be systematic biases correlated with 
morphological type. Since we have assumed an exponential profile, a
galaxy with an $r^{1/4}$ profile will not be well represented by our
model (see Section~\ref{sec_profiles}).

In summary, although the APM measurements have relatively large
uncertainties, they are adequate to classify galaxies into broad
surface brightness categories.

\subsection{Comparison with other SB measurements of UKST survey plates}

Morshidi-Esslinger et al.  (1999, hereafter MDS) made a survey of
a substantial number of UKST survey plates in the south galactic
cap in an attempt to identify LSBGs with large angular size. They
used the APM to measure profiles of nearly 3000 LSBGs and
estimated the central surface brightness by extrapolating the best
fit exponential profile. Their SB estimates are based on the same
original plates as our data, and they use the same scanning
machine. They also make the same assumption of exponential galaxy
profiles, so the MDS survey is not a completely independent data
set. Nevertheless it provides a useful comparison for our
estimates.

There are two main differences between our measurements and those
of MDS. First, they fitted exponential profiles to up to 8
isophotal areas $A_{0},A_{1},...,A_{7}$ -- see the details in
Maddox et al.  (1990a) and MDS (1999) -- to solve for the
($p_{0}$, $r_{0}$) parameters, whereas we used only $A_{0}$ to
calculate these values. The second difference is in the
calibration procedure. MDS's calibration of SB depends on
zero-points for total magnitudes from other works for fields where
this is available, and uses the magnitudes of stars in plate
overlaps for fields with no external calibration. Our calibration
depends mainly on matching $\mu_0$ directly for the galaxies plate
overlaps, with an overall calibration from external photometry.

\begin{figure}
\centerline{\hbox{\psfig{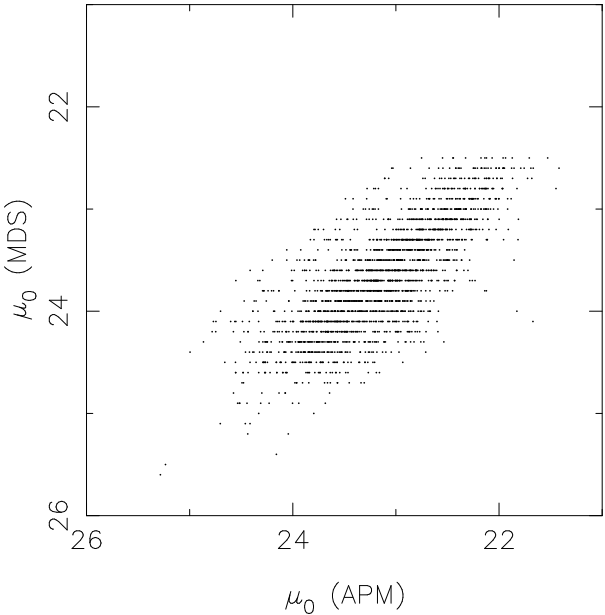}}}
\caption{A comparison of central surface brightnesses of MDS and
this paper. The dispersion is $\sigma = 0.35$ mag arcsec$^{-2}$.
To be consistent with MDS(1999), we have applied a correction to
our $\mu_{0}$ to account for the ellipticities of galaxies.
\label{fig_mds} }
\end{figure}

\begin{figure}
\centerline{\hbox{\psfig{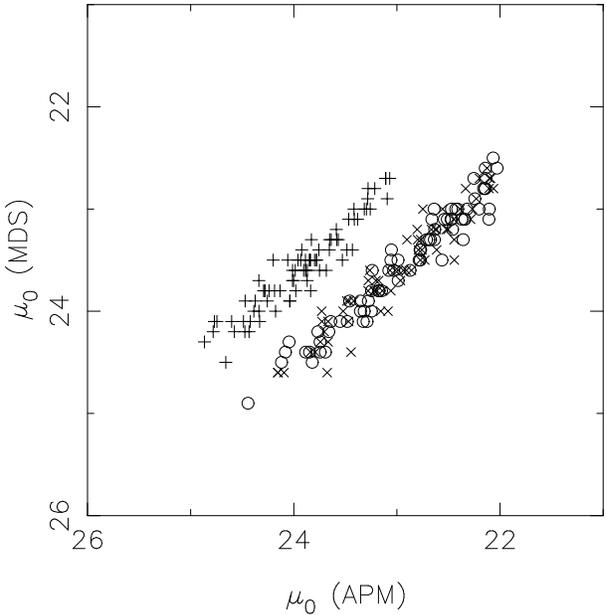}}}
\caption{The same as figure ~\ref{fig_mds} but showing data from
only three independent plates which have more than 65 MDS galaxies
on each. The three fields are: F157, which has an {\em rms}
scatter of $0.16$ mag arcsec$^{-2}$, is plotted with diagonal
crosses; F358, which has an {\em rms} scatter of $ 0.15$ mag
arcsec$^{-2}$, is plotted with vertical/horizontal crosses; F482,
which has an {\em rms} scatter of $\sigma = 0.14$ mag
arcsec$^{-2}$, is plotted with open circles. \label{fig_mds2}}
\end{figure}

Figure~\ref{fig_mds} compares our estimates of $\mu_{0}$ with 2185
MDS galaxies that can be identified in our sample. The
unidentified galaxies are mainly classified as `merged' images in
the APM galaxy survey, and thus not included in our catalogue.
There is a very good correlation between the two estimates, but
the scatter is rather large, with an {\em rms} of 0.35 mag. To
investigate the cause of this dispersion, we have examined data
from three fields (F157, F358 and F482) having more than 65 MDS
galaxies in each, and have plotted them with different symbols in
Figure~\ref{fig_mds2}. It is clear that the {\em rms} scatter for
each plate is very small, only 0.16, 0.15 and 0.14 respectively.
These scatters are slightly smaller than the typical {\em rms}
uncertainties estimated from the APM plate overlaps. So, even if
we assume that the MDS measurements are correct, and all of the
scatter is due to the approximation involved in our simple method
of estimating $\mu_0$, the errors introduced are smaller than the
observational errors associated with measuring different UKST
plates.

These three plates were deliberately chosen to lie at the extremes
of the distribution plotted in Figure~\ref{fig_mds}, and it is
clear that the zero-points of these three plates are very
different. The zero-point for field F358, is almost 1 mag away
from other two plates.  If our zero-point was in error by 1 mag,
it would be clearly visible as a discrepant point in
Figure~\ref{fig_meanmu0}. Also the area covered by field F358
would show up as a large overdensity in Figure~\ref{fig_skymu}(a),
and an underdensity in Figure~\ref{fig_skymu}(c). None of these
tests suggests there is any significant error in our zero-point
for this plate, and so we believe the MDS zero-point is in error.
Assuming that the {\em rms} error on our measurements is 0.16 mag
as discussed in Section~\ref{sec_errors}, this suggests that the
MDS zero-points have an {\em rms} of 0.31 mag, and so F358 would
be a 3 sigma deviation, which is not unexpected in a sample of
around 100 plates.

MDS estimate that the uncertainty in their calibration is about
0.1 mag for plates with external calibration, and 0.2 for plates
where they have used overlaps. There are two possible sources for
the larger errors that we find. First, they do not apply any sort
of vignetting or desensitization correction. Since their
zero-points rely on edge matching and small-area CCD calibration
data, this will introduce an error in the plate zero-points with
an {\em rms} of order 0.1-0.2 mag. Second, MDS use the magnitudes
of faint stars in the plate overlaps to estimate the zero-points
on plates with no external calibration. Even faint stars tend to
have a higher SB than galaxies, and this is likely to introduce a
further bias. These two effects together might account for the
observed 0.31 mag {\em rms} difference. This comparison shows that
calibrating the zero-points of photographic data is a difficult
problem that is still without a satisfactory solution.

\begin{figure*}
\hbox{\psfig{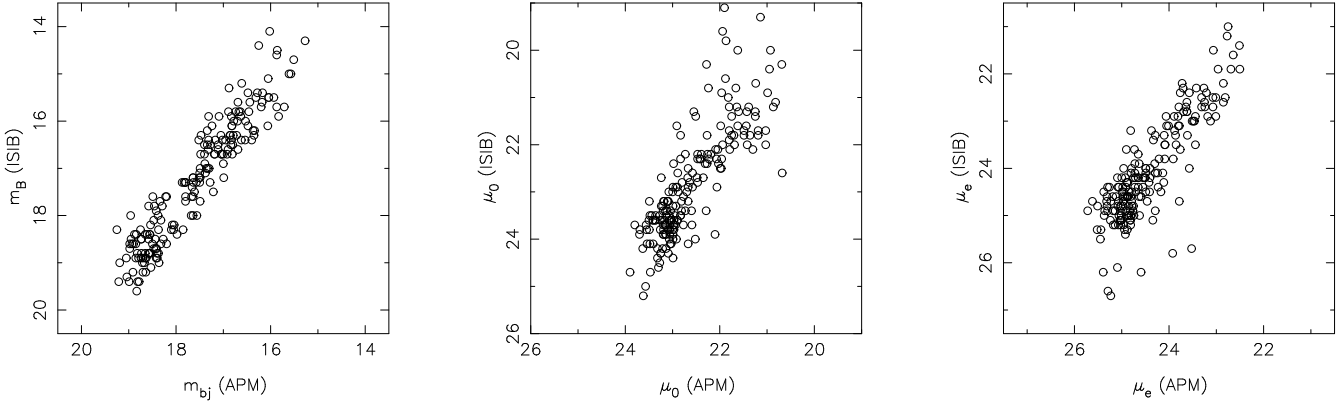}}
\caption{Comparison of total magnitude, central surface brightness
and effective surface brightness of ISIB and this paper. Note that
all data from ISIB are in the B band, while our APM results are in
the $b_J$ photographic band. 
 \label{fig isib}
}
\end{figure*}

Impey et al. (1996, hereafter ISIB) scanned 24 UKST survey plates to
search for LSBGs with $\mu_0>22$~mag arcsec$^{-2}$. Some of these
plates are covered by the APM data used in this paper, and we find 202
galaxies in common with their data.  Plots comparing the our
magnitudes and surface brightnesses with the ISIB measurements are
shown in Figure~\ref{fig isib}. There is a good correlation between
the ISIB estimates and ours, but there are several differences in
their analysis and definition of parameters which preclude a
quantitative comparison with their data.

\section{Discussion}
\label{sec_discussion}

\subsection{The Distribution of Surface Brightnesses}
\label{sec_sb}

\begin{figure}
\centerline{\hbox{\psfig{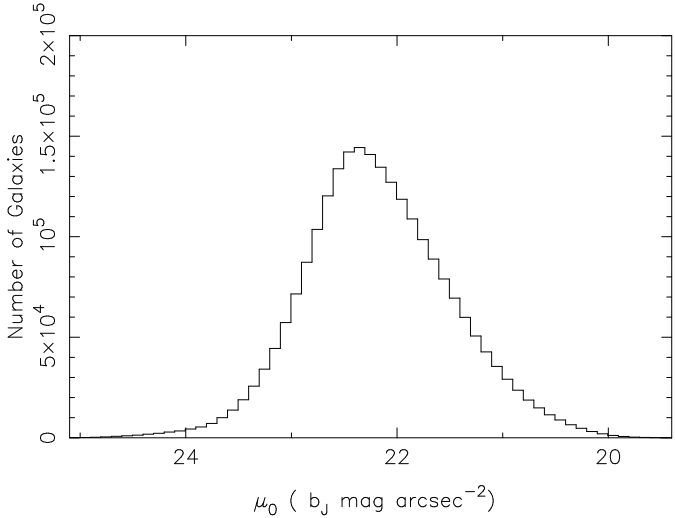}}}
\caption{The frequency distribution of the apparent SB, $\mu_0$, for
all galaxies in the survey.
\label{fig_mu0hist}
}
\end{figure}

 Figure~\ref{fig_mu0hist} shows a histogram of the frequency
distribution of $\mu_0$. The distribution is centred on $\mu_0 =
22.2 $ mag arcsec$^{-2}$, and extends to $\mu_0 \simeq 20$ mag
arcsec$^{-2}$ at the high SB end, and to $\mu_0 \simeq 24.5$ mag
arcsec$^{-2}$ at the low SB end.  The star-galaxy separation
criteria (Maddox et al. 1990a) mean that some very high SB images
will be rejected from the galaxy sample because their profile is
so similar to stars. However, the rapid drop in the apparent
number of high SB galaxies will be due to their small isophotal
areas and a real drop in the space density of high SB galaxies (as
shown by Cross et al. 2001).  The lack of very low SB galaxies is
less clearly a reflection of the true distribution of galaxy SBs,
since a galaxy must have at least 4 square arcseconds brighter
than 25 mag arcsec$^{-2}$ to be selected in the APM galaxy sample.
The distribution is further constrained by the 20.5 mag limit on the
Gaussian profile magnitudes which will also bias against low SB
galaxies. 
Cross et al. (2001) have used the 2dF Galaxy Redshift Survey to
estimate the distribution of intrinsic surface brightnesses, and
conclude that, although the APM selection criteria do exclude low
SB galaxies, the missing galaxies would not contribute
significantly to the overall luminosity density. We will consider
how our selection criteria affect the observed distribution in a
future paper.

It is clear from Figure~\ref{fig_mu0hist} that the $\mu_0$ subsamples
plotted in Figure~\ref{fig_skymu}(a) and \ref{fig_skymu}(c) represent the
galaxies at the extremes of SB in the APM sample. This means that the
distribution of galaxies in the sub-samples is very sensitive to any
residual errors in the uniformity of the $\mu_0$ measurements, and so
the apparent uniformity of the galaxies in Figure~\ref{fig_skymu}(a) and
\ref{fig_skymu}(c) show that our corrections have given us a reliable
and uniform set of $\mu_0$ measurements.

\subsection{The profiles of APM bright galaxies}
\label{sec_profiles}

In this Section, we use galaxies from the APM Bright Galaxy
Catalogue (Loveday 1996; hereafter APMBGC) to investigate how the
profile of a galaxy depends on its morphological type, and test if
the assumption of an exponential profile is valid. The galaxies in
the APMBGC have $b_J < 16.44$ and have a relatively large angular
size ($r_{\rm iso}$) and large $\sigma^{2}$. Although $\sigma^{2}$
is not very sensitive to changes in the profile of galaxies, and
the observational error is quite large, this parameter does
contain useful information about the SB profile. Furthermore, the
APMBGC galaxies have been morphologically classified by visual
inspection of the UKS plates and we have analyzed the distribution
of $\sigma^{2}$ separately for both early and late type galaxies.

We calculated the ($p_{0}$, $r_{0}$) of each galaxy, assuming an
exponential SB profile as in Section~\ref{sec measurement}. Since
we also know the apparent ellipticity $e$ for each galaxy (from
the density weighted second moments), and the values of the
detection threshold and saturation of the plate, we can simulate
the observed APM density for each pixel. Any other observational
parameters can then be estimated from this exponential model
image. In particular, we have estimated the value of $\sigma^{2}$,
and compare this to the actual observed value. Obviously, if the
original image is exactly an exponential profile, and there are no
other observational errors, then the observed value, $\sigma_{\rm
obs}$ will be exactly equal to that from the exponential model
$\sigma_{\rm exp}$, so the ratio $\sigma_{\rm obs}/\sigma_{\rm
exp}$ should be exactly unity. To test the sensitivity of $\sigma$
to variations in the SB profile, we created simulated images that
follow the $r^{1/4}$ law profile (eq.~\ref{eq_pr4}), and analysed
them using the assumption of an exponential profile. In this case
we found that $\sigma_{\rm obs}/\sigma_{\rm exp} \simeq 0.9$, with
a small variation according to their real central surface
brightness and apparent ellipticity. The same analysis for
Gaussian profiles gave a ratio of $\sigma_{\rm obs}/\sigma_{\rm
exp} \simeq 1.1$.

\begin{figure}

\centerline{\hbox{\psfig{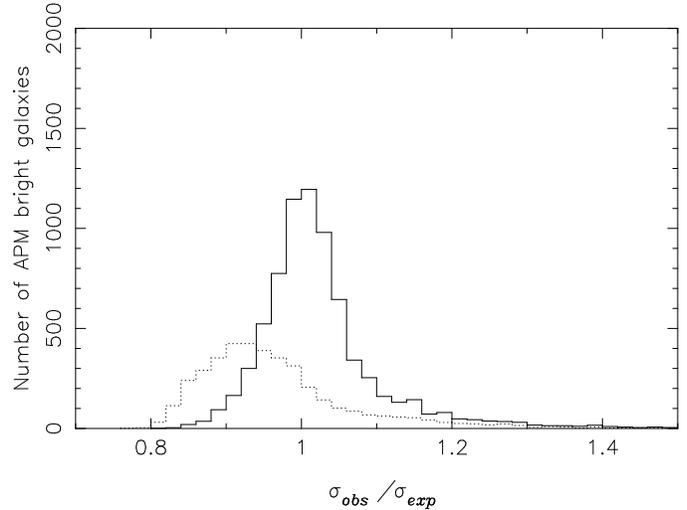}}}
\caption{The frequency distribution of $\sigma_{\rm
obs}/\sigma_{\rm exp}$ values for APM bright galaxies. The dotted
line shows E and S0 galaxies; the solid line shows Spiral and Irr
. An exponential profile would have $\sigma_{\rm obs}/\sigma_{\rm
exp}=1.0$, an $r^{1/4}$ law profile would have $\sigma_{\rm
obs}/\sigma_{\rm exp}=0.9$, and a Gaussian profile would have
$\sigma_{\rm obs}/\sigma_{\rm exp}=1.1$. \label{fig_sig2} }

\end{figure}

In Figure~\ref{fig_sig2}, we plot the frequency distribution of
$\sigma_{\rm obs}/\sigma_{\rm exp}$ for early and late galaxies from
the APMBGC.  There is a large dispersion in $\sigma_{\rm
obs}/\sigma_{\rm exp}$ , even for only late type galaxies.
Nevertheless, the peak of the distribution is very close to 1.0, which
implies that the exponential profile is a reasonable approximation for
most galaxies  (77\% of the sample has $0.9< \sigma_{\rm
obs}/\sigma_{\rm exp} < 1.1$).  For early type galaxies (E and S0),
the dispersion is larger than for late type galaxies, and most of them
have smaller $\sigma_{\rm obs}/\sigma_{\rm exp}$ values that are more
consistent with an $r^{1/4}$ law profile.  This shows that,
although the APM measurements span a rather limited range of SB, the
$\sigma$ parameter can distinguish between early and late-type
galaxies.

For fainter APM galaxies, the observational uncertainty on
$\sigma^{2}$ will be much larger, and the small differences
between the $r^{1/4}$, exponential and Gaussian assumptions mean
that $\sigma^{2}$ is not effective in constraining the SB profile.

\section{Summary}
\label{sec_summary}

We have described a simple method to estimate the central surface
brightness, $\mu_0$, of galaxies, allowing for saturation and
isophotal effects, and applied it to all the galaxies in the APM
galaxy survey. We have corrected the $\mu_0$ estimates for the effects
of vignetting and differential desensitization over the field of each
UKS plate. We have also used the plate overlap areas to match the
individual plate measurements, and so provide homogeneous $\mu_0$
estimates for all the APM galaxies.  Internal and external checks of
the uncertainties in $\mu_0$ suggest that the uncertainty for an
individual measurement is about 0.16 mag.

For very bright APM galaxies, the second radial moment of the SB
profile, $\sigma^2$, shows that late type galaxies are consistent
with an exponential profile, whereas early type galaxies have a
higher mean SB, consistent with an $r^{1/4}$ law.

This uniform set of SB estimates for the APM galaxies will be used
in later papers to select samples of high and low SB galaxies and analyze
differences between their distributions.\\

\bigskip

\section*{Acknowledgements}

Z.S. would like to thank all members of the astronomy group in
Nottingham for their hospitality during his stay in Nottingham.
Z.S. acknowledges support from grant NKBRSF G19990754 and the
National Nature Sciences Foundation of China. This work was also
partly supported by PPARC. 
We thank the referee for useful comments on the original manuscript. 
The comparison with CCD observations used the NASA Goddard 
Space Flight Center HDF-S public data recorded by J.P. Gardner 
and P. Palunas at the Cerro Tololo Inter-American Observatory, 
a division of the National Optical Astronomy Observatories. 
Further comparisons used the Anglo-Australian Observatory 
Hubble Deep Field-South public data, recorded by K. Glazebrook, 
R. Abraham and C. Tinney on the Anglo-Australian Telescope.

\def\aj {AJ}
\def\apj {ApJ}
\def\aap {A\&A}
\def\aaps {A\&AS}
\def\ajs {AJS}
\def\apjs {ApJS}
\def\mn {MNRAS}
\def\apjl {Ap. J. Let.}
\def\pasp {PASP}
\def\SE {}

\appendix

\label{lastpage}


\begin{thebibliography}{99}


\bibitem{B1} Bertin E., 1998, \SE {\it SExtractor User's Guide,}
         Institut d'Astro\-physique de Paris
\bibitem{B2} Bertin E., Arnouts S., 1996, \aaps, 117, 396
\bibitem{B3} Blair M., Gilmore G., 1982, \pasp, 94, 742
\bibitem{C1} Cawson M. G. M., Kibblewhite E. J., Disney M. J.,
         Phillipps S., 1987, \mn, 224, 557
\bibitem{C2} C\^{o}t\'e S., Broadhurst T., Loveday J., Kolind S.,
             1999, ASP Conference Series, 170,
             {\it The Low Surface Brightness Universe.}
             IAU Col. 171, 307
\bibitem{C3} Cross N., Driver S. P., Couch W. J.,  et al.,
         2001, \mn, 324, 825
\bibitem{D1}Dawe J. A., Metcalfe N., 1982, {\it Proc. astr. Soc. Aust.},
         4, 466
\bibitem{d1}de Jong R. S., 1996, \aaps, 118, 557
\bibitem{d2}de Jong R. S., van der Kruit P. C., 1994, \aaps, 106, 451
\bibitem{H1}Heraudeau P., Simien F., 1996, \aaps, 118, 111
\bibitem{I1} Impey C. D., Sprayberry D.,  Irwin M. J.,  Bothun, G. D.,
         1996, \apjs, 105, 209
\bibitem{J1}Jansen R. A., Franx M., Fabricant D., Caldwell N., 2000,
          \apjs, 126, 271
\bibitem{L1} Lambas D. G., Maddox S. J., Loveday J. 1992, \mn, 258, 404
\bibitem{L2}Lauberts A., Valentijn E. A., 1989, {\em The Surface Photometry
          Catalogue of the ESO-Uppsala Galaxies}. European Southern
          Observatory, Garching
\bibitem{L3} Loveday J., 1996, \mn, 287, 1025
\bibitem{M1} Maddox S. J., 1988. {\it PhD thesis}, University of Cambridge.
\bibitem{M2} Maddox S. J., Efstathiou G., Sutherland W. J.,
         Loveday J., 1990a,  \mn, 243. 692
\bibitem{M3} Maddox S. J., Efstathiou G., Sutherland W. J., 1996,
         \mn, 283, 1227
\bibitem{M4} Maddox S. J., Efstathiou G., Sutherland W. J.,
         1990b, \mn, 246, 433
\bibitem{M5} Morshidi-Esslinger Z. B.,  Davies J. I. \& Smith R. M.,
         1999, \mn, 304, 297
\bibitem{P1}Palunas P., Collins N. R., Gardner J. P., Hill R. S., Malumuth
            E. M.,  Smette A., Teplitz H. I., Williger G. M., Woodgate
            B. E., 2000,  \apj, 541, 61
\bibitem{S1} Schlegel D., Finkbeiner D., Davis, M., 1998,
            \apj, 500, 525
\bibitem{s2} Sprayberry D., Impey C., Irwin M., 1996, \apj,
                463, 535
\bibitem{T1}Teplitz H. I., Hill R. S., Malumuth E. M., Collins N. R., Gardner
        J. P., Palunas P., Woodgate B. E., 2001, \apj, 548, 127
\bibitem{W1} Wittman D. M., Tyson J. A., Bernstein G. M., Lee R. W.,
        dell'Antonio I. P., Fischer P., Smith D. R., Blouke M. M., 1998.
        Proc. SPIE Vol. 3355, 626



\end{thebibliography}
\end{document}